\documentclass[aps,prb,reprint]{revtex4-1}
\usepackage[dvips]{graphicx}
\usepackage{amsmath}
\usepackage{color}

\begin{document}

\title{Exciton diffusion in semiconducting single-wall carbon nanotubes studied by transient absorption microscopy}

\author{Brian~A.~Ruzicka,$^1$ Rui Wang,$^1$ Jessica Lohrman,$^2$ Shenqiang Ren,$^{2*}$ and Hui~Zhao$^{1}$}\email{huizhao@ku.edu; shenqiang@ku.edu}

\affiliation{$^1$Department of Physics and Astronomy, The University of Kansas, Lawrence, Kansas 66045, USA\\
 $^2$Department of Chemistry, The University of Kansas, Lawrence, Kansas 66045, USA}
\date{\today}

\begin{abstract}
Spatiotemporal dynamics of excitons in isolated semiconducting single-walled carbon nanotubes are studied using transient absorption microscopy. Differential reflection and transmission of an 810-nm probe pulse after excitation by a 750-nm pump pulse are measured. We observe a bi-exponentially decaying signal with a fast time constant of 0.66~ps and a slower time constant of 2.8~ps. Both constants are independent of the pump fluence. By spatially and temporally resolving the differential reflection, we are able to observe a diffusion of excitons, and measure a diffusion coefficient of 200$\pm$10~cm$^2$/s at room temperature and 300$\pm$10~cm$^2$/s at lower temperatures of 10~K and 150~K.
\end{abstract}

\maketitle

Single-walled carbon nanotubes (SWNTs) have attracted considerable attention for the last two decades.\cite{n363603} Their unique mechanical, electrical, and optical properties have made them an attractive candidate for many applications.\cite{s2701179,n381678,n386377,n39349,s287622} Owing to the strong Coulomb interaction in this one-dimensional structure, the interaction between electrons and light is dominated by exciton effects. Hence, understanding the exciton dynamics in SWNTs is important for many optoelectronic applications. Photoemission\cite{l845002} and transient absorption\cite{apl81975,l90057404,am15534,apl821458,l92017403,l97207401,l96027401,b73075403} measurements on samples of SWNT bundles, where semiconducting and metallic tubes are entangled together, have shown ultrafast exciton dynamics characterized by an energy relaxation time of about 0.1~ps and an exciton lifetime of about 1~ps. Other transient absorption studies on samples of isolated or individual tubes revealed that these fast dynamics are induced by fast transfer of excitons from semiconducting to metallic tubes. When such channels are eliminated in isolated tubes, the energy relaxation can take several picoseconds,\cite{l93017403,b71125427,l92117402,b71115444,b80205411} and exciton lifetimes of several 10~ps\cite{b71125427,l92117402} to several 100~ps\cite{l93017403,b71115444} have been measured. Time-resolved photoluminescence measurements have also shown decay times of several 10~ps.\cite{l92177401,l95197401,nl5511,b71033402,b71115426,l97257401} Other aspects of exciton dynamics have also been studied, including exciton-exciton annihilation,\cite{l94157402} exciton dephasing,\cite{nl83936} exciton-phonon interactions,\cite{np2515,nl83102,l102127401,b80245428} multiple exciton generation,\cite{apl92233105,nl102381} intraexciton transition,\cite{l104177401} and exciton nonlinearities.\cite{l94047404,b80201405}

In contrast to these extensive studies in time and energy domains, the exciton dynamics in real space, i.e. exciton diffusion along the tubes, has been rarely studied. Since the size of excitons (about 2~nm\cite{np554}) is much smaller than the tube length, exciton diffusion plays an essential role in excitation energy transfer in many optoelectronic applications. From a fundamental point of view, studies of exciton diffusion can provide valuable information on microscopic interactions between excitons and their environment in nanotubes.

A direct measurement of exciton diffusion is challenging, since electrical techniques that are typically used for transport studies are less effective on excitons that are electrically neutral. Recently, several attempts have been made to deduce the exciton diffusion coefficient in SWNT samples. However, the results differ by orders of magnitude. For example, a depolarization effect observed in early transient absorption measurements was attributed to the exciton diffusion in curved SWNTs, suggesting a diffusion coefficient of 120~cm$^2$/s.\cite{l92017403} By measuring exciton lifetime as a function of exciton density, one can determine the exciton-exciton annihilation rate.\cite{l94157402,b73115432} This can be modeled to deduce the diffusion coefficient. However, measurements using transient absorption and time-resolved photoluminescence techniques have yielded rather different results: the former found very small diffusion coefficients of 0.1~cm$^2$/s\cite{np554} to 4~cm$^2$/s;\cite{acsnano59898} while the latter gave a value of about 90~cm$^2$/s.\cite{b70241403} Other transient absorption measurements were interpreted by considering exciton diffusion to quenching sites like defects or tube ends, and resulted in diffusion coefficients of about 10~cm$^2$/s.\cite{acsnano47161,jpcc1113831,b74041405}

In addition to these ultrafast studies in the time domain, stepwise quenching of exciton luminescence by single-molecule reactions was used to deduce exciton diffusion lengths of 60~nm\cite{s3161465} to 200~nm,\cite{jpcl12189} and simulations of near-field microscopy measurements of photoluminescence quenching at the tube ends gave diffusion lengths of 100 - 200~nm.\cite{pssb2462683,jpcc1144353} Furthermore, by modeling the power dependence of photoluminescence with different tube lengths, a diffusion length of 610~nm was deduced.\cite{l104247402} Since in these studies the exciton lifetime was not measured, one has to assume a certain lifetime in order to estimate the diffusion coefficient. This and the rather large range of the diffusion lengths deduced has resulted in a large range of the estimated diffusion coefficients, including 0.4~cm$^2$/s,\cite{s3161465} 2.5-10~cm$^2$/s,\cite{pssb2462683} and 44~cm$^2$/s.\cite{l104247402}

The discrepancy in these studies could be attributed to the lack of a model-independent technique to directly measure the diffusion coefficient. Here we show that a transient absorption microscopy with high spatial and temporal resolution can be used to solve this issue by time-resolving the exciton transport in real space. By using a tightly focused femtosecond pulse, excitons with a thin spatial distribution are excited in a film of isolated semiconducting SWNTs wrapped by poly(3-hexylthiophene) (P3HT). Diffusion of these excitons is directly monitored by time-resolving the broadening of the density profile, which is achieved by measuring the transient absorption of a time-delayed and spatially scanned probe pulse. The diffusion coefficient is directly obtained from the rate of increase in the profile area. We obtain a diffusion coefficient of $200\pm10$ cm$^2$/s at room temperature, and $300\pm10$ cm$^2$/s at lower temperatures.

\begin{figure}
\centering
\includegraphics[width=8.5cm]{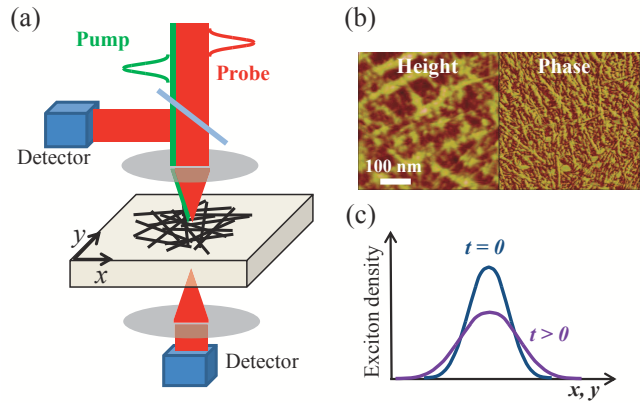}
\caption{(a) Experimental configuration: A tightly focused pump pulse excites excitons in a thin film of isolated and randomly oriented semiconducting SWNTs. A time-delayed and spatially scanned probe pulse detects the exciton density as a function of time and space via differential reflection and differential transmission. (b) Atomic force microscopy image of the thin film sample. (c) After excitation, the excitons diffuse to lower density regions along the nanotubes, causing a broadening in the spatial profile over time.}
\label{fig:configuration}
\end{figure}

High-purity semiconducting SWNTs with nominal diameter in the range of 1.2 - 1.7~nm and length in the range of 300~nm - 5~$\mu$m (IsoNanotube-S from NanoIntegris, 98\% purity) were used in this study. The nanotubes are in conjunction with regioregular P3HT (99\% head-tail coupling, average molecular weight MW $\approx$ 50,000). The P3HT passivated SWNT nanohybrids were prepared in a 10 mg/mL 1,2-dichlorobenzene solution containing P3HT and SWNTs (3 wt\%). The synthesis of core-shell nanohybrids was promoted by adding the non-solvent acetonitrile to favor P3HT aggregation onto the SWNT surface. Additional sonication steps were used to further enhance the P3HT backbone assembly to passivate the SWNT surface, in order to avoid the tube bundling. Transmission electron microscopy image confirmed the formation of core/shell SWNTs coated by P3HT shell layers.\cite{nl115316} From the absorption spectrum, the second exciton transition fall in the range of 800-1200 nm, as predicted by the Kataura plot.\cite{nl115316} A thin film of the passivated SWNTs was spin-coated on a glass slide at 1000 rpm for one minute and dried at 100$^{\mathrm{o}}$C for 10 minutes. Fig.~\ref{fig:configuration}(b) shows an atomic force microscopy image of the sample. The thin film has an optical density of 0.36 at 810 nm.

The experimental geometry is illustrated in Fig.~\ref{fig:configuration}(a). A pump pulse with 750-nm wavelength and about 200 fs temporal width is focused to the sample to a spot size of about 2~$\mu$m (full width at half maximum, FWHM). It excites excitons at the second energy level. Initially, the spatial profile of the excitons is thin, as shown in Fig.~\ref{fig:configuration}(c). After a short time, excitons diffuse in each SWNT, causing the overall density profile to expand. By solving the diffusion equation with an initial gaussian distribution, one has
\begin{equation}
w^{2}(t)=w^2(t_0)+16\ln(2)D(t-t_{0}),
\label{eq:w2vst}
\end{equation}
where $w(t_0)$ is the width (FWHM) of the profile at time $t_0$  and $D$ is the diffusion coefficient of excitons.\cite{b79115321,b82195414,l89097401,apl801391} Hence, by measuring the evolution of the exciton density profile, we can directly determine $D$.

We probe the exciton density by measuring the transient absorption of a 200-fs and 810-nm probe pulse that is also tightly focused to about 2~$\mu$m by the same lens [Fig.~\ref{fig:configuration}(a)]. The reflected probe from the sample is directed to a detector by using a beamsplitter. The differential reflection $\Delta R/R_{0}=(R-R_0)/R_0$, i.e. the normalized change in reflection of the probe pulse caused by the pump-injected excitons, is measured by modulating the intensity of the pump pulse with an optical chopper at 2.1~kHz and using a lock-in amplifier. Here, $R$ and $R_0$ are the reflection with and without the presence of the pump pulse, respectively. The transmitted probe is collimated and sent to another detector in order to simultaneously measure the differential transmission, defined similarly as $\Delta T/T_{0}=(T-T_0)/T_0$. In these measurements, balanced detection technique is used to improve the signal-to-noise ratio.\cite{l96246601}

\begin{figure}
\centering
\includegraphics[width=8.5cm]{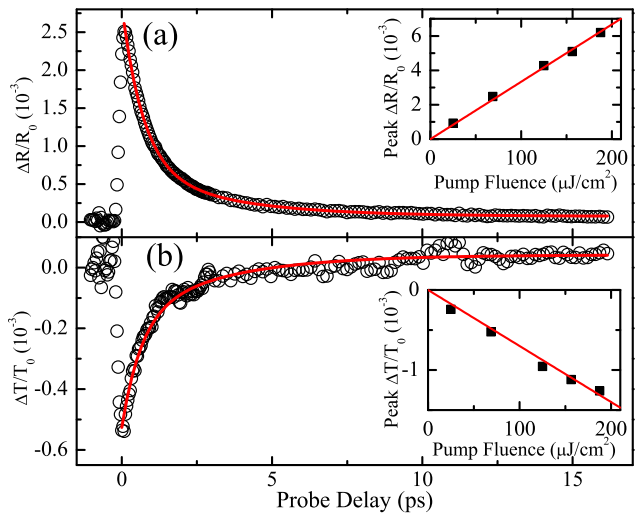}
\caption{Differential reflection (a) and transmission (b) measured with a pump fluence of 70~$\mu$J/cm$^2$ and with the pump and probe spots overlapped on the sample. The red curve in (a) is a bi-exponential fit to the data. The red curve in (b) uses the same time constant from (b). The insets show the peak differential reflection and transmission as a function of pump fluence.}
\label{fig:DRvsF}
\end{figure}

We first measure the time-resolved differential reflection and transmission with the pump and the probe laser spots overlapped on the sample. The probe and the pump pulses are both linearly polarized, along the $x$ and $y$ directions, respectively. The sample is at room temperature. Figure~\ref{fig:DRvsF} shows the results with a pump pulse energy fluence of 70~$\mu$J/cm$^2$. The differential reflection and transmission are positive and negative, respectively. Previous studies have shown that the transient absorption can be positive or negative, i.e. the pump-injected excitons can increase (photoinduced absorption) or decrease (photo-bleaching) the absorption of the probe pulse, depending on the detuning of the probe photon energy with respect to the excitonic transitions.\cite{l90057404,l92017403,l94157402} The negative differential transmission suggests that we are in the photoinduced absorption regime. The differential reflection curve can be fit very well with bi-exponential functions, as shown as the red curve in Fig.~\ref{fig:DRvsF}(a). The two time constants from this fit are 0.68 and 3.0 ps, respectively. The simultaneously measured differential transmission can be described with the same two time constants, as shown as the red curve in Fig.~\ref{fig:DRvsF}(b). We repeat the measurement with different pump fluences. The magnitude of the peak signals increases linearly with the fluence, as shown in the insets of Fig.~\ref{fig:DRvsF}. Since the injected exciton density is proportional to the pump fluence, we conclude that both signals are proportional to the exciton density. Furthermore, we find that the shape of both curves remains the same as we vary the pump fluence. By averaging the time constants deduced from all the measured differential reflection curves, we deduce a short time constant of 0.66$\pm$0.02 ps and a long time constant of 2.8$\pm$0.2 ps. The long time constant is consistent with the energy relaxation time measured previously in isolated semiconducting SWNTs.\cite{l93017403,b71125427,l92117402,b71115444,b80205411} This indicates that the photoinduced absorption we observed originates from the excitons in the second energy level. We attribute the short time constant to thermalization of the excitons. In this process, the distribution of excitons evolves from the initial gaussian distribution (determined by the spectrum of the pump pulse) to a thermal distribution via exciton-exciton scattering.

Since the differential reflection and transmission signals are both proportional to the exciton density and the decay times are independent of the exciton density, the time evolution of the exciton density profile can be monitored by these signals, regardless of the mechanism of the photoinduced absorption. This allows us to directly monitor the diffusion process of the excitons in real space. Since the differential reflection signal is stronger and less sensitive to the sample nonuniformity, compared to the differential transmission, we use it to study the exciton diffusion.

\begin{figure}
\centering
\includegraphics[width=8.5cm]{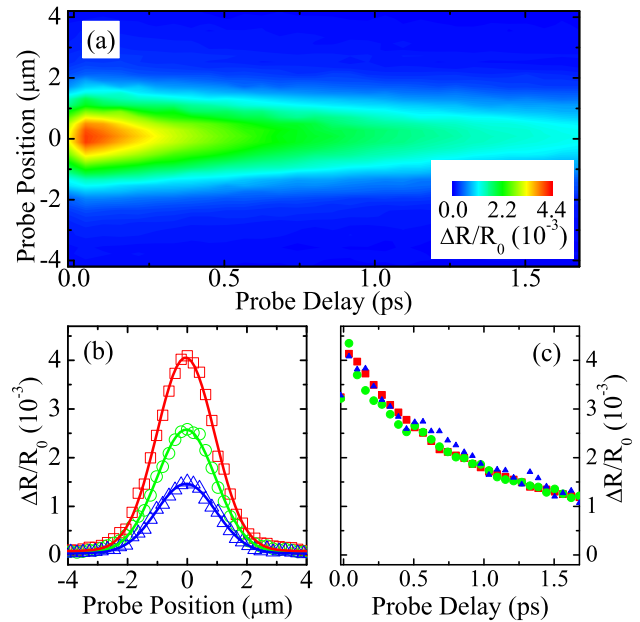}
\caption{(a) Differential reflection as a function of probe position and probe delay, measured with a pump fluence of 125~$\mu$J/cm$^2$ and with the sample at room temperature. (b) Differential reflection as a function of probe position for probe delays of 0~ps (red squares), 0.8~ps (green circles), and 1.6~ps (blue triangles). (c) Differential reflection as a function of probe delay for probe positions of 0~$\mu$m (red squares), 1.2~$\mu$m (green circles), and 2.4 $\mu$m (blue triangles), respectively. In order to compare the decay rates, the latter two have been scaled, by multiplying by factors of 2.3 and 12, respectively, to match the first.  }
\label{fig:DRvsSpace}
\end{figure}

\begin{figure}
\centering
\includegraphics[width=8.5cm]{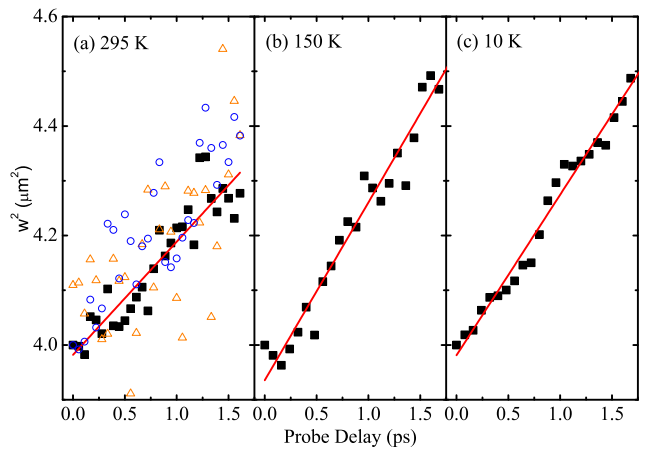}
\caption{(a) Solid squares: squared width of the differential reflection profile as a function of probe delay, obtained from Gaussian fits of the profiles shown in Fig.~3; Open triangles: with a lower pump fluence of 60~$\mu$m/cm$^2$; Open circles: with a pump polarization parallel to the probe and a pump fluence of 125~$\mu$J/cm$^2$. (b) and (c): same as solid squares in (a), but with sample temperatures of 150 K and 10 K, respectively.}
\label{fig:diffusion}
\end{figure}

We measure the differential reflection signal as a function of the probe delay and the probe spot location with respect to the pump spot location, with a pump fluence of 125 $\mu$J/cm$^2$. The results are plotted in Fig.~\ref{fig:DRvsSpace}(a). Here $0~\mu$m is defined where the centers of the probe and pump spots are overlapped. At each probe delay, the signal has a gaussian spatial profile. Figure~\ref{fig:DRvsSpace}(b) shows a few examples of the profiles. We fit each profile with a gaussian function [solid curves in Fig.~\ref{fig:DRvsSpace}(b)] to deduce the width (FWHM), $w$. The squared width is plotted in Fig.~\ref{fig:diffusion}(a) (solid squares) as a function of the probe delay. As expected according to Eq.~\ref{eq:w2vst}, the squared width increases linearly with time. A fit (red line) gives a diffusion coefficient of $200\pm10$~cm$^2$/s. Using the exciton lifetime in the excited state of $\tau_e$ = 2.8 ps, we can also deduce a diffusion length of $\sqrt{D \tau_e}$ = 240 nm for the excitons in the excited state.

Fugure \ref{fig:DRvsSpace}(a) also shows the decay of signal at each probe position, which is consistent with the results shown in Fig.~\ref{fig:DRvsF}. To see how the dynamics vary with probe positions, we plot three time scans with fixed probe positions of  0~$\mu$m (red squares), 1.2~$\mu$m (green circles), and 2.4 $\mu$m (blue triangles), respectively, in Fig.~\ref{fig:DRvsSpace}(c). The latter two curves are normalized with respect to the first one for better comparison. Clearly, all three curves have similar delay rates. The time evolution of the signal at each probe position is governed by energy relaxation and recombination of excitons at that position, which causes an overall decay of the signal, and transport of excitons from and to adjacent positions. Apparently, the later does not contribute significantly to the dynamics. This is, however, expected since the deduced diffusion length of 240 nm is only a small fraction of the laser spots used in this study.  It also confirms that the decay constants deduced from Fig.~\ref{fig:DRvsF} are not significantly influenced by the diffusion.

The broadening of the exciton density profile provides unambiguous evidence of the exciton diffusion in nanotubes. Owing to the direct spatial resolution of the excitonic dynamics, this procedure of measuring the diffusion coefficient does not rely on sophisticated models and assumptions. It is also not influenced by other processes and experimental conditions. For example, the exciton relaxation and recombination do not influence the result, since these processes only change the height, but not the width, of the profile. Similarly, the measurement does not rely on our attribution of the two decay constants of 0.66 and 2.8 ps to thermalization and energy relaxation of excitons, since these processes only influence the height. The finite probe spot size does not influence the result, either. Because both spots are gaussian, the convolution of the actual profile with the probe spot only adds a constant to the measured squared width. It does not change the slope.\cite{apl92112104} In the measurements, we took time scans at each probe location, instead of spatial scans at each probe delay. Therefore, drift of the sample out of the focal plane would not cause an apparent broadening.\cite{b83235306,b86045406} It would have caused asymmetric gaussian profiles, which was not seen in Fig.~\ref{fig:DRvsSpace}. Finally, since most nanotubes are 2 - 3 $\mu$m long, about 10 times longer than the diffusion length, recombination of excitons at tube ends is expected to play a minor role in the dynamics.

We repeat the measurement with a reduced pump fluence of 60~$\mu$J/cm$^2$, and hence a lower exciton density. The result is also plotted in Fig.~\ref{fig:diffusion}(a) (open triangles). Due to a lower signal level, the measured width shows a larger uncertainty. However, the slope, and hence the diffusion coefficient, is consistent with the 125-$\mu$J/cm$^2$ result. Since it is known that light absorption depends strongly on the direction of the tube with respect to the light polarization, we also repeat the measurement with an $x$-polarized pump pulse of 125 $\mu$J/cm$^2$, i.e. parallel to the probe polarization. The open circles in Fig.~\ref{fig:diffusion}(a) show the measured squared width. The larger uncertainty compared to the cross-polarization measurement (squares) is due to a higher noise level caused by the pump beam, since in this configuration we could not use a polarizer in front of the detector to block the pump. However, the slope is very similar to the cross-polarization measurement.

To study the exciton diffusion as a function of sample temperature, we repeat the measurement with sample temperatures of 150 and 10~K, respectively. The squared width as a function of probe delay for these two temperatures are shown in Fig. \ref{fig:diffusion}(b) and (c). At the lower temperatures, we observe a diffusion coefficient of approximately $300\pm10$~cm$^2$/s, the same for both. While this value is slightly higher, we do note that as the temperature changes from 10 to 293 K we are unable to maintain the same sample position and therefore the difference may be due to sample nonuniformity. Nevertheless, we can conclude that the diffusion coefficient of excitons in SWNTs does not change significantly with temperature. 

Diffusion coefficient of charge carriers is related to carrier mobility, $\mu$, by Einstein's relation, $D=\mu k_{B}T/e$, where $k_B$, $T$, and $e$ are Boltzmann constant, temperature, and elementary charge, respectively. The mobility is proportional to the mean free time, which is determined by scattering rates. When transport is limited by phonon scattering, the mobility increases with decreasing temperature. Previous studies\cite{l95146805} have shown that in the temperature range of 80 - 250 K, $\mu \sim T^{-1}$. Such a result suggests that the diffusion coefficient is independent of temperature. Since excitons are electrically neutral, their phonon scattering, and hence diffusion, can be different from charge carriers. However, our conclusion that exciton diffusion coefficient has no strong temperature dependence is similar to behavior of charge carriers.

The experimental procedure is based on broadening of the exciton density profile in the sample plane. In this two-dimensional diffusion process, excitons change their motion direction randomly when encountered a scattering event. For exciton diffusion in carbon nanotubes, the excitons can only move along the tube direction. However, since the tubes are randomly oriented, they mimic a two dimensional random diffusion process. Hence, we do not expect the measured quantities of diffusion coefficient to be much different from the actual diffusion coefficients in carbon nanotubes. To confirm this, we used a standard Monte Carlo simulation to directly compare the two conditions.\cite{b67035306,l94137402} First, we simulate the two dimensional diffusion. Excitons are generated with randomly selected positions that fulfill a Gaussian distribution and with randomly selected speeds that fulfill a Boltzmann distribution. The direction of velocity of each exciton is purely random in the sample plane. Each exciton then follows a series of free-flight / scattering events. The duration of each free flight is determined randomly according to a mean free time. For our purpose, it is sufficient to only include elastic scattering. Hence, after each scattering, the speed of the exciton is kept unchanged, while the direction of velocity is randomly selected. We verified that the diffusion coefficient obtained from the broadening of the profile is consistent the expected value based on the temperature and mean free time. We then simulate the diffusion in the thin film of randomly oriented carbon nanotubes by modifying the program such that at each scattering event, the velocity direction is either unchanged or reversed. The random orientation of the carbon nanotubes is incorporated in the simulation since the direction of initial velocity of each exciton is random. Based on this simulation, the diffusion coefficients in the two situations are within 10\%.

In summary, we have performed a spatially resolved transient absorption study of exciton diffusion in isolated semiconducting SWNTs wrapped by P3HT. Spatiotemporal dynamics of excitons injected by a tightly focused pump pulse are studied by measuring differential reflection and differential transmission of a time-delayed and spatially scanned probe pulse. We observe a bi-exponentially decaying signal with a fast time constant of 0.66 ps and a slower time constant of 2.8 ps. Both constants are independent of the pump fluence. The squared width of the exciton density profile increases linearly with time, as expected for a diffusion process. We measured a diffusion coefficient of 200$\pm$10~cm$^2$/s at room temperature, which is independent of the pump fluence.  We additionally investigated the diffusion coefficient at temperatures of 10 and 150 K and found diffusion coefficients of approximately 300$\pm$10~cm$^2$/s at both. This direct measurement of exciton diffusion coefficient in semiconducting SWNTs can help to solve the controversy on this important aspect of exciton dynamics. The results provide valuable information that can be used to understand other aspects of exciton dynamics in carbon nanotubes.

BAR, RW, and HZ acknowledge support from the US National Science Foundation under Awards No. DMR-0954486 and No. EPS-0903806, and matching support from the State of Kansas through Kansas Technology Enterprise Corporation. S.R. thanks the University of Kansas for its startup financial support and acknowledge a subcontract support from a Department of Energy award (DESC0005448).


%

\end{document}